# Electrical transport and thermoelectric properties of silver nanoparticles

Vikash Sharma and Gunadhor Singh Okram[#]

UGC-DAE Consortium for Scientific Research, University Campus, Khandwa Road, Indore 452001, Madhya Pradesh, India.

Email: [#]okram@csr.res.in

**Abstract** Debye temperature ($\theta_D$) decrease, residual resistivity ($\rho_0$) increase manifesting inapplicability of Bloch-Grüneisen (BG) theorem and electron-phonon coupling constant ($\alpha_{e-ph}$) increase as crystallite size decreases have been found from electrical resistivity ($\rho$) in temperature (T) range 5 K to 300 K of well-characterized Ag nanoparticles (NPs) synthesized with oleylamine (OA), trioctylphosphine (TOP) and/ polyvinylpyrrolidone (PVP) with Scherrer sizes ranging from 15.1 nm to 33.4 nm. Notably, ~ 36 % reduction in $\theta_D$ in 15.1 nm compared to bulk Ag is found. Remarkably, usual phonon drag peak found in Seebeck coefficient (S) for bulk Ag turned into dips or phonon drag minima (PDM) in these NPs that gradually gets suppressed and shifted towards lower T with decrease in crystallite size in OA and TOP-induced NPs. Contrastingly, it appears at higher T in TOP-induced NPs. A broad hump between 125 K to 215 K, a slope change near 270 K in resistivity and an additional dip-like feature near 172 K in S are seen in OA-PVP-induced NPs with different shapes. They are attributed to spatial confinement of electrons and phonons, varying barrier heights, different charge-transfer mechanisms among metal NPs and surfactant/s, enhanced disorders (grain boundaries (GBs), increase in fraction of surface atoms, surfactant matrix and other defects), leading to modifications in their overall electron and phonon interactions. Finally, their thermoelectric power factor has also been assessed.

1. ## Introduction

Silver nanostructures are of great interest due to their numerous potential applications in plasmonics, nanoelectronics and optoelectronics[1,2,3,4,5,6,7] due their fascinating optical properties, high electrical and thermal conductivity[1,2,3,8] originated from surface plasmon resonance, high electronic density of states (DOS), surface states and surface plasmons[4,9]. In particular, thermoelectric (TE) properties, generally related to conversion of heat to electricity, provide the unique information about the charge transport that is different from simple electrical transport properties[10]. Many metals like Ag, Au, Cu and Li have positive sign of Seebeck coefficients (S) as oppose to the sign of charge carriers or Hall coefficient that is expected. This can only be explained using energy dependent conductivity $\sigma(E)$ or electron lifetime ($\tau(E)$) and its derivate with energy. It cannot obtain from Drude's formula, $\sigma = \frac{ne^2\tau}{m}$, since DOS or $n$ and $\tau$ has inverse relation and hence $\sigma$ and $(\frac{1}{m})$ has same energy dependence[10]. The electrical and TE properties are highly dependent on crystallite size, size distribution, shape and surface-bound molecules[4,7,11,12,13,14]. The overall relaxation rates and electron-phonon scattering increases as crystallite size decreases due to spatial confinement of electrons, phonons and their additional scattering with impurities (defects, GBs, surfaces and surfactant) and hence influences the transport properties[11,12]. The surface-bound molecules of surfactants are of special interest as they may lead to strong modification in electrical and TE transport properties[9,13,15,16]. In fact, while distinguishing the role of each effect is somehow difficult, their cumulative effects will influence the overall transport behaviour[3,8,17,18,19,20,21,22].

Nevertheless, metals have poor value of TE figure of merit, $ZT = \frac{S^2}{\rho\kappa}T$ due to smaller value of S and high thermal conductivity ($\kappa$) and low $\rho$. However, small metallic NPs are found to be beneficial to improve the ZT using their inclusion in parent compound through electron filtering and enhance phonon scattering. For example, enhancement in electrical conductivity as well as slightly increase in S by inclusion of Ag NPs and large reduction in thermal conductivity by scatterings of phonons is achieved by grain size reduction[23,24]. The enhancement in S for metallic surface has been achieved using surface state enhancement and manipulating their electronic properties. Rettenberger *et al.* have reported significantly large value of S = -45±5 µV/K for Ag (111) surface[25], and Maksymovych *et al.* found significant enhancement in S ~ -90 µV/K in Ag (111) terraces due to surface state, which are one to two orders of magnitude larger compared to single-atom step sites and surface-supported NPs on Ag (111) surface[4] or its bulk counterpart[26,27]. It is therefore motivating to study the electrical and TE properties in metallic nanostructures to improve upon the fundamental understanding to utilize them for particular application[2,3,8,7,17,18,19,20,28].

Considerable efforts have been put forward to understand the role of size and surface effects on electrical transport properties using $\rho$ in T range 5 K to 300 K for single nanowire, nanowire arrays, nanowire networks and NPs of Ag[2,3,8,18,20]. S for bulk Ag and its alloys in T range of 5 K to 300 K have been extensively studied[26,27,29,30]. However, the effects of size and surface on S for Ag nanostructures are poorly understood with only one report for Ag NPs in T range of 77 K to 300 K[19] and a report on a single Ag nanowire between 150 K to 300 K[8]. Moreover,

TE properties for Ag (111) surface at only 300K[25] and over T range of 5 K to 400 K[4] have also been reported. However, there has not been any experimental report on S and $\rho$ for Ag NPs with varying crystallite sizes and surfactant/s in T range 5 K to 300 K.

We report here studies on the influence of crystallite size and surfactant/s on $\rho$ and S in T range 5 K to 300 K. These properties are found to be enormously modified compared to its bulk counterpart. The residual resistance $\rho_0$ and electron-phonon coupling constant $\alpha_{e-ph}$ are increased while the Debye temperature $\theta_D$ is decreased with decrease in crystallite size. Anomalously, the usual phonon drag peak for bulk Ag evolves into a gradually suppressing scenario turning into a phonon drag dip or PDM with associated gradual shift towards lower T with decrease in crystallite size for samples prepared in combination of OA and TOP. However, it appeared at higher T in NPs prepared with TOP only. Furthermore, a broad hump-like feature between 125 K to 215 K and a slope change near 270 K in resistivity data appear, and an additional dip-like feature near 172 K in S is seen in the Ag NPs prepared with OA and PVP, which appears to be associated with presence of different shapes of NPs. These results are explained using spatial confinement of electrons and phonons, varying barrier heights and different charge transfer mechanisms between metal NPs and surfactant/s and enhanced disorders, and hence modification in their overall electrons and phonons relaxation processes.

Sample synthesis conditions and crystallite size calculated from Scherrer relation are listed in table A1 (in Appendix); details of characterization of these Ag NPs have been reported previously[31]. The Seebeck coefficient S and resistivity $\rho$ were measured of cold-pressed NP samples in T range of 5 K to 300 K in a specially designed commercially available Dewar using differential direct current and a home-made four-point probe setups[32,33].

## 2. Results and Discussion
### 2.1 Evolution of electrical resistivity with crystallite size

The electrical resistivity $\rho$ of Ag1, Ag2, Ag3, Ag4, Ag5 and Ag6 shows (figure 1) the systematic increase towards higher values compared to the bulk Ag with clear metallic nature in each case. In order to be able to appreciate this trend better, plots for Ag1 and Ag2 and bulk are shown in figure 1, inset (a). To see how the resistivity is increasing with crystallite size, it is shown as absolute value (figure 1, inset (b)) and in their ratios with respect to the corresponding values of the bulk at 5 K and 300 K (figure 1, inset (c)). It is therefore, clear that even though the resistivity in absolute value looks small, their ratio with respect to its bulk is significantly enhanced as the crystallite size decreases; it is largest (4615) for Ag4. This trend is related to the samples Ag1 to Ag4, prepared in OA as TOP increases. However, in Ag5, prepared in 8 ml TOP only, resistivity at 300 K is larger than that of Ag3 or Ag6 but less than that of Ag4 and its slope is interestingly the largest of all the samples. On the other hand, for Ag6, prepared in 12 ml TOP only, its value at 300 K is smaller than that of Ag5 but larger than that of Ag3 with its slope nearly those of Ag1-Ag3 but much smaller than that of Ag5 or Ag4. The trend for Ag7, prepared in OA-PVP, is curiously completely different (figure 1, inset (d)): it is not so linear compared to Ag1-Ag6, even though it shows, in general, metallic behaviour with a broad hump between 125 K to 215 K peaking near 165 K. In addition, the change in slope can be seen near 270 K.

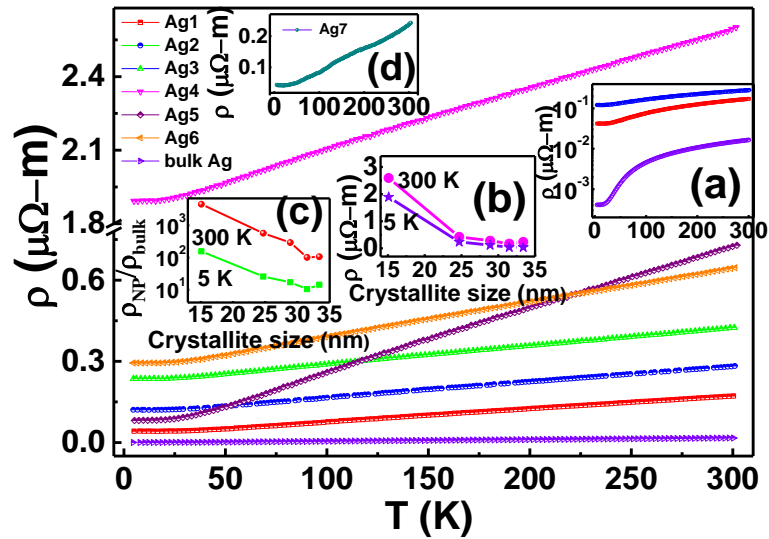

**Figure 1** Resistivity of Ag1, Ag2, Ag3, Ag4, Ag5, Ag6 and bulk Ag as a function of temperature. Insets: (a) Ag1, Ag2 and bulk Ag to understand the features better, (b) resistivity at 5 K and 300 K versus crystallite size, (c) resistivity ratio at 5 K and 300 K (with respect to the corresponding values of the bulk) versus crystallite size and (d) resistivity of Ag7 as a function of temperature.

## 2.2 Influence of crystallite size on electron-phonon interactions, residual resistivity ratio and Debye temperature

In order to understand more clearly, resistivity curves of Ag1, Ag2, Ag3, Ag4, Ag5 and Ag6 were fitted with eq. A1 for n=5 with $\theta_D$ and $\alpha_{e-ph}$ are adjustable fit parameters (figure A1). The optimized relative fit error defined as $(\rho_{measured} - \rho_{fit})/\rho_{measured} \times 100\%$ found to be less than ± 1 % in Ag1, Ag2, Ag3, Ag4, Ag6 and ± 1.9 % in Ag5 is shown in inset of each graph. The obtained parameters $\theta_D$ and $\alpha_{e-ph}$ are listed in table A2. The metallic behaviour of $\rho$ of all samples and its decrease with decrease in T is attributed to increase in mobility and reduction in electron-phonon (acoustic) scattering. This is in fact so due to decrease in number of phonons, and they are frozen at sufficient low T, say T < 30 K for Ag1. Below this, only scattering of charge carriers with impurities remains, whereas electron-phonon scattering vanishes and $\rho$ becomes almost independent of T. However, $\rho$ can be upturned at low T as reported for 5.3 nm Ag NPs[20] which is not the case in the present data ruling out the absence of localization effects or drastic disorders. To study the effect of crystallite size and the effect of surfactant/s, we discuss $\rho$ separately for NPs prepared with combination of OA and TOP (OA-TOP) i.e. Ag1, Ag2, Ag3, Ag4, and prepared in pure TOP i. e. Ag5 and Ag6.

The residual resistance ratio (RRR) i.e. ratio of $\rho$ at 300 K to 5 K ($\rho_{300}/\rho_5$) determines the quality or disorder present in the sample. It is found to be decreasing with decrease in crystallite size in samples prepared in OA-TOP as well as in pure TOP (figure 2 (a), table A2), consistent with earlier report[20]. However, these values of RRR is significantly smaller than that obtained from bulk Ag (table A2) that is consistent with increase in $\rho_0$ in the NPs. Enhanced GBs, increase in fraction of surface atoms and other defects including enhance thickness of surfactant matrix with decrease in crystallite size have critical roles in the electrical transport of the NPs. Even though the RRR of these NPs is significantly deviated from linearity due to dominant scattering of electrons with GBs and surfactant, it does not decay exponentially as in Ag nanowires[17,18]. In nanowires, only surface scattering is dominated and fraction of atoms on crystallite surface increases in a smooth scaling law[34].

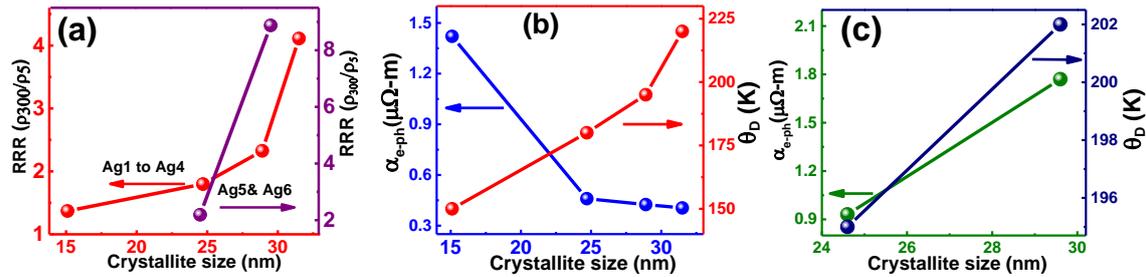

**Figure 2** (a) Residual resistivity ratio for Ag1, Ag2, Ag3, Ag4 (red curve) and Ag5 and Ag6 (violet curve), electron-phonon coupling constant and Debye temperature of (b) Ag1, Ag2, Ag3, Ag4 and (c) Ag5 and Ag6 as a function of crystallite size.

The electron-phonon coupling constant $\alpha_{e-ph}$ significantly increases for OA-TOP-induced NPs (figure 2 (b) and table A2) while it decreases for TOP-only induced NPs (figure 2 (c)) with decrease in crystallite size that can also be readily seen from slope of $\rho$ (figures 1 and S2). The $\rho_{e-ph}$ gives rise to the major resistance for clean metal can only be described by electron-phonon scattering. However, for NP systems, $\rho_{e-ph}$ is due to many other scattering mechanisms such as GBs and surfactant disorders[3,17]. The increase in $\alpha_{e-ph}$ manifests the increase in electron-phonon scattering due to spatial confinement of electrons and phonons with simultaneous increase in scattering of electrons with GBs and decrease in probability of tunnelling or transmission through GBs as a consequence of increase in thickness of surfactant matrix, as evident from the large difference in the crystallite size and TEM size for Ag5 in particular (Table A1). The larger value of $\alpha_{e-ph}$ in Ag5, compared to smaller size NP Ag6, seems to be due to less disorders that led to decrease in scattering of electrons and phonons with them, which is consistent with significantly larger value of RRR in Ag5 than that of Ag6 (figure 2 (a)). Furthermore, it is worth to note that electron-phonon interaction or $\alpha_{e-ph}$ is not just dependent upon crystallite size, but it is also considerably affected by size distribution or agglomeration, lattice defects and other surface properties[35,36]. Jain *et al.* [35] showed that enhanced agglomeration of NPs leads to increase in $\alpha_{e-ph}$, which is consistent with highest degree of agglomeration in Ag5 compared to others (see figure 5d, ref[31].).

The $\theta_D$ systematically decreases with decrease in crystallite size in both OA-TOP-induced and TOP-induced NPs (figure 2 (b, c)), which is significantly reduced by ~ 36 % in 15.1 nm sample Ag4 compared to its bulk value ~ 234 K (table A2), and consistent with earlier report[3]. This signifies that softening in acoustic phonons with reduction in crystallite size, consistent with earlier reports[18,28]. This is likely to happen due to the weakening of the surface energy and enhanced fraction of surface atoms or enhanced contribution of surface-dangling bonds with decrease in crystallite size. This leads to decrease in elastic modulus and hence decrease in $\theta_D$. The temperature coefficient of resistivity (TCR), defined as $1/\rho(d\rho/dT)$, rises more rapidly in bigger size NPs compared to smaller with decrease in T, and then start to decrease (figure A3). A peak each near 50 K, 46 K, 42 K, 35 K, 47 K and 44 K for Ag1, Ag2, Ag3, Ag4, Ag5 and Ag6 is found, respectively, that shift at lower T with decrease in crystallite size as in Ag1-Ag4 (figure 3 (a), left axis) or in Ag5 and Ag6 (figure 3 (a), right axis). TCR found to be positive for all samples is expected for metal, although it turns to negative value at low T due to fluctuations in data points. It shows decreasing value with decrease in crystallite size (figure 3 (a)), and attains maximum value at around 1.98 % K$^{-1}$ near 47 K for Ag5.

The influence of surfactant on electrical transport properties is interesting[13], mainly due to the alteration in surface properties, GBs, and other chemical disorders including surfactant. This is evident in Ag3 and Ag6 which have different values of RRR, $\theta_D$, $\alpha_{e-ph}$ and TCR (figure 2 (b, c), 3 (a)) even though they have nearly same crystallite size. The $\theta_D$ is sensitive to many surface properties such as surface energy, bond strength and contribution of bonds, which may change in NPs synthesized using different surfactants due to the different degree of disorders[14], agglomeration or interparticle interaction[35] and crystallite size distribution that in turn affect the relaxation rates of electrons and phonons[35] and scattering of phonons and electrons with surface and hence electrical transport properties[9,13]. This is clear from the different values of RRR (figure 2(a)), $\theta_D$, $\alpha_{e-ph}$ and TCR (figure 2 (b, c), 3 (a)). Such trend is not exactly followed in Ag7 as these fits could not be made with it because of the broad hump and slope change that is attributed to the presence of different shaped-NPs with broad size distribution[35] (see figure 5 (i) in ref[31]).

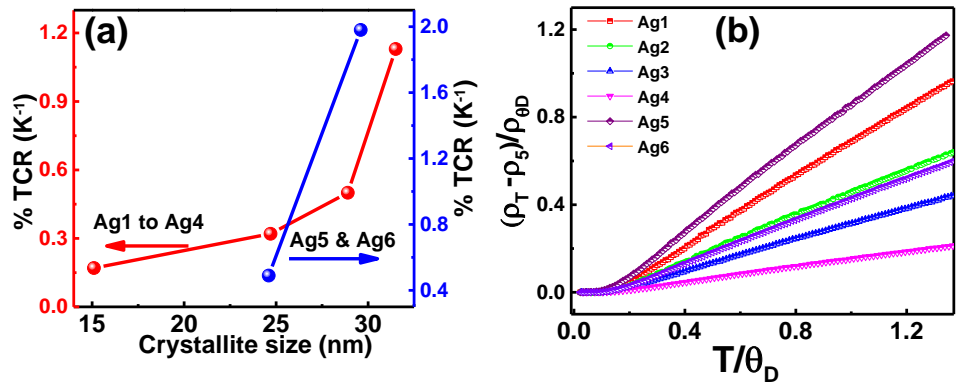

**Figure 3** (a) temperature coefficient of resistance of Ag1, Ag2, Ag3 and Ag4 (red curve) and Ag5 and Ag6 (blue curve) and (b) scaling law for Ag1, Ag2, Ag3, Ag4, Ag5 and Ag6.

**2.3 Deviation from Bloch-Gruneisen theorem**

The validity of Bloch-Gruneisen (BG) theorem can be assessed using one-parameter scaling law, wherein all curves plotted between $\frac{\rho - \rho_5}{\rho_{\theta_D}}$ versus $\frac{T}{\theta_D}$ collapse into one curve[18]. According to the BG theorem, basic electron–acoustic-phonon interaction as well as the simple Debye phonon spectrum, i.e. phonon density of states $D(\omega) \propto \omega^2$, where $\omega$ is phonon frequency remains unchanged with decrease in size of material[28]. It is clearly seen that all curves are not collapsed in one curve (figure 3 (b)). This means that BG theorem is not applicable for these NPs in line with change in $\alpha_{e-ph}$ with reduction in crystallite size, which is mainly due to confinement of electrons and phonons, increasing contribution of GBs and other defects including surfactant/s. As a consequence, overall relaxation time of phonons, electrons and electron-phonon interactions are modified. Bid *et al.* however observed that this theorem holds for Ag nanowires of diameter down to 15 nm without considerable variation in $\alpha_{e-ph}$ with decrease in dimension of wire nm[18]. The contrasting results observed here compared to that of Bid *et al.* indicate the dominant role of the type of the surfactants used.

**2.4 Influence of crystallite size on thermopower**

The trends in S for Ag1, Ag2, Ag3, Ag4, Ag5, Ag6 and Ag7 are shown in figure 4. As T raises, S of Ag1 decreases with the minimum near 63 K, attributed as phonon drag minimum (PDM), above which it increases approximately linearly with T. The other samples show more or less similar features as Ag1 with minima at 63 K, 59 K, 54 K and 47 K with S values 0.57 µV/K, 0.53 µV/K, 0.49 µV/K and 0.42 µV/K for Ag1, Ag2, Ag3 and Ag4, respectively (figure 4 (a)). PDM normally appears due to phonon drag effects that dominate at low T regime ($\theta_D/10 < T < \theta_D/5$), is gradually suppressed and shifted towards lower T with decrease in crystallite size (figure 4 (a)), which are similar to those reported earlier in other metal NPs[17,21]. In fact, the decrease in $\theta_D$ can be related to the decrease in phonon frequency and hence decrease in wavevector of longitudinal phonons ($u_l$). Under free electron model and relaxation time ($\tau$) approximation, with $\tau$ falling as size decreases due to enhanced scattering, reduction in $S_g$ requires less decline in $u_l$ in the present situation of decreasing $\theta_D$ (i.e. T) as it has to balance each other according to eq. A11. The physical significance of shift to lower (higher) PDM temperature with shrinkage in crystallite size is related to larger (smaller) distortion in lattice that in turn gives rise to decrease (increase) in mean free path of the phonons. The linear increase in S above PDM for Ag1 and Ag2 is attributed to electron diffusion contribution ($S_d$) that dominates at high T regime[19]. However, $S_d$ increases for Ag3 and Ag4 with T are not that linear as those of Ag1 and Ag2. The hump-like feature near 125 K in both Ag3 and Ag4, and a dip near 220 K for Ag4 seen (figure 4 (a)) might be indicative of modifications in diffusive transport in smaller size NPs[17,21] and dominant role of TOP. Notably, $S_d$ in high T regime increase with reduction in crystallite size (figure 4 (a), inset) is consistent with earlier reports[17,21] while below 230 K, it decreases with reduction in crystallite size (figure 4(a)). Interestingly, sign of S is positive in all the samples in the T range of 5 K to 300 K like that of bulk Ag (figure 4(a)) and as reported[26,27,29,30]. This is opposite to the sign of actual majority (negative) charge carriers, in line with other metals[10,17]. The positive sign of S cannot be explained using constant time relaxation, but dependency of energy on electrical conductivity ($\sigma(E)$) i.e. $\frac{d\sigma(E)}{dE}$ (see eq. A3)[10].

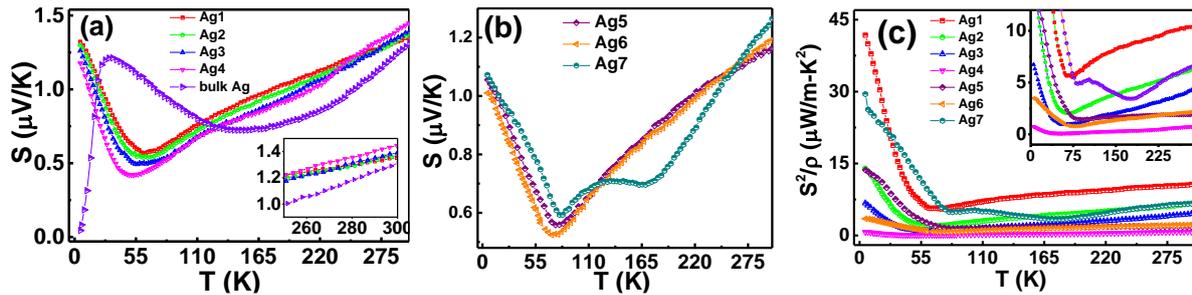

**Figure 4** Seebeck coefficient of (a) Ag1, Ag2, Ag3 and Ag4 and (b) Ag5, Ag6 and Ag7, and (c) their power factor. Insets: (a) higher temperature region of S and (c) lower power factor region.

The S of Ag5 and Ag6 exhibit approximately similar trends like Ag1. The PDM appear near 75 K and 73 K for Ag5 and Ag6 with values around 0.56 µV/K and 0.53 µV/K, respectively (figure 4 (b)). The PDM are supressed and shifted towards lower T as crystallite size drops. However, the PDM are found to be at significantly higher T in these TOP-induced Ag5 and Ag6 NPs, compared to OA-TOP-induced Ag1, Ag2, Ag3 and Ag4 NPs. Interestingly, S of Ag7 exhibits two PDM-like features near 80 K with value 0.59 µV/K and 172 K with value 0.70 µV/K (figure 4 (b)). The former is assigned as PDM as found in other samples, while second is more likely to be due to change in diffusion contribution that may be associated with very random triangular, rectangular, hexagonal and spherical shapes of NPs and broad size distribution[31]. Moreover, this dip-like feature is completely different and more pronounced than found in Ag4 around 220 K due to increase in diffusive contribution in Ag4. Such type of unusual broad hump between 125 K to 215 K, peaking near 165 K in resistivity of Ag7 is also seen (figure 3(b)). The $S_d$ for Ag7 is smaller in T range 125 K to 220 K compared to Ag5 and Ag6 and exhibits a dip-like feature near 172 K. The dip temperature at 172 K is very high as per the standard rule for normal metallic samples as $\Theta_D/10 < T < \Theta_D/5$ and therefore, not expected to be another phonon drag effect. However, it seems to be related to diffusion contributions of electrons of two types of ensembles of NPs which might have their distinct contributions different from each other having different S slopes between approximately 80-100K and above 175 K, as a consequence of different scattering mechanisms of electrons with phonons, impurities or GBs[37]. While shape of NPs is found to be less important to affect the electron-phonon relaxation[36], different shapes and their broad size distribution along with interactions with surfactant molecules can change density of GBs, periodicity and symmetry of lattice resulting in modified inelastic electron and phonon scattering. In addition, different shapes of NPs may have modified shape of Fermi surface and electron distribution that may lead to change in asymmetric scattering barrier or particle hole asymmetry around Fermi level, and hence affect the electron and hole transport across the barrier that leads to modification in $S_d$[4,38].

### 2.4.1 General discussion on thermopower of silver on alloying or nanostructuring

S in T range of 5 K to 300 K for bulk Ag and its alloy has been studied in details[26,27,29,30] in which there is a gradual suppression in phonon drag peak and shift in peak position towards lower T, and decrease in diffusion contribution with increase in atomic % of secondary element in Ag as an alloy. Crisp *et al.*[26] found that phonon drag peak turns to a dip with negative value for $Au_{0.85}Ag_{0.15}$ and Lee *et al.*[27] found that $S_d$ turns negative from positive on alloying Ag with In, Sn, Sb, Ga and Ge. Schroeder *et al.*[29] observed a crossover from positive to negative value and a dip-like feature near 15 K for Ag-Pd alloy with 0.53 % and 1.28 % Pd, wherein unlike the former, the latter shows an upturn at low temperature that was concluded to be not due to phonon drag effect. Cusack *et al.*[30] also reported similar features. There is however reported inconsistency in the value of phonon drag peak and S at 300 K for bulk Ag in these reports[26,27,29,30]. Hence, S is highly sensitive to defects and imperfections in the sample.

Reports on S for Ag nanostructures are rare. They are on Ag NPs in T range 77 K to 300 K[19] and single Ag nanowire between 150 K to 300 K[8]. Zhu *et al.*[19] showed increased $S_d$ with a valley-like feature near 110 K for bulk Ag that gets gradually shifted to lower T with decrease in crystallite size ( ~95 K for 11 nm NPs). They suggestively concluded that phonon drag peak should shift to lower T with decrease in crystallite size on the basis of valley position, without proper information below 77 K. We in contrast observe minima (PDM), not peak, in S for Ag NPs due to phonon drag effects. Therefore, assumption of Zhu *et al.*[19] in shift in the valley position towards lower T has perhaps been ruled out, wherein phonon drag peak in bulk turn out to be PDM, with its position shifting to lower temperature as the size decreases (figure 4 (a)). Furthermore, presently observed $S_d$ value increasing only at high T, say above 250 K for Ag4, not in whole T range, and its decreasing value below 230 K for Ag1 to Ag4 with reduction in crystallite size is in contrast to their report[19] but is, to some extent, consistent with its reported smaller value in Ag alloys[26,27,29,30] compared to bulk Ag. Comparison of Kojda *et al.*[8] of their observed S value (seemingly associated with Pt contact), less than 0 to 5.7 µV/K in T ~150 K to 300 K for Ag nanowire, as nearly equal, to those of S ~0.4 to 1.3 µV/K in T~10 K – 300 K for bulk Ag appears surprising[29,30] since even bulk Pt/ Pt wire has large negative value in this temperature range[32,33].

To explain these features, one has to understand the effect of different scattering mechanisms on diffusive $S_d$ and phonon drag $S_g$ contributions to the total thermopower, S. The $S_d$ calculation assumes to happen in thermal equilibrium of the diffusion of electrons from hot end to cold end and is found to vary as $\left(\frac{T}{E_F}\right)$ in free electron approximation (see eq. A5 and A6)[39]. However, thermal equilibrium is lifted in the presence of thermal gradient, contributing non-equilibrium phonons in S. Thus, $S_g$ is originated from non-equilibrium phonons and delivers excessive momenta to the electrons through electron-phonon interactions that leads to extra electrical current in the direction of heat flow, i.e. phonons drag electrons[40]. It increases as $\left(\frac{T}{\theta_D}\right)^3$ in low T limit as T decreases due to decrease in phonon–phonon scattering (see eq. A10 or A11)[17,21,41]. Thus, long wavelength phonons drag electrons to appear as a peak at low T[17,21,41]. It is qualitatively clear that $S_g$ will be maximal when electron-phonon scattering dominates over other scattering mechanisms such as electron-impurity, phonon-impurity and phonon-phonon scattering.

In the context of the present NPs that have Scherrer/ crystallite size (where the periodicity is maintained) less than mean free path of electrons of bulk Ag (52 nm[42]), when the crystallite size falls below the mean free path of electrons, finite size (related to electronic structure) and surface effects (related to impurities) dominate. Therefore, understanding how the finite size and surface effects influence electrical transport and TE properties in NPs is rather intriguing. Their thermopower depends on many scattering mechanisms such as scattering of electrons and phonons with impurities (defects, GBs, surfaces and surfactant matrix) along with electron-phonon interactions, with the difficulty to individually separate their influences, and hence cumulative effects prevail as the actual transport behaviour. The presence of other scattering, electron/ phonon confinement and decreased $\tau$ for electron-phonon scattering appear to manifest as increased $\alpha_{e-ph}$ as size falls since, for example, $\alpha_{e-ph}$ and $\tau$ are inversely related[36]. This leads to modification in S. The finite size effects may modify DOS that in turn are proportional to the number of electrons involved to form the band structure[34]. It is therefore expected that DOS or charge carrier concentration should decrease around Fermi level (i.e. decrease in Fermi energy ($E_F$)) with diminution in crystallite size that should surge $S_d$ since it is inversely related to $E_F$ (see eq. A7, A9 and A10)[17,21].

Furthermore, it is noted that $S_g$ is critically dependent on electron-phonon interactions while long mean free path phonons mainly contribute to phonon drag compared to those that carry heat[40] and their transports are affected due to scattering of impurities and spatial confinement[11,12]. Also, spatial confinement of electrons and phonons result in modification in their velocity, mean free path and increase in relaxation rates as the crystallite size decreases[11,18,43]. The observed increase in $\alpha_{e-ph}$ that indicates decrease in electron-phonon relaxation time due to their inverse relation[36], reflect other obstacles like GBs, surface and surfactant matrix, whereby phonons cannot perhaps drag electrons beyond the GBs. Therefore, presently observed suppression in $S_g$ with decrease in crystallite size in Ag NPs is more likely to be due to[11,12]: (i) decrease in mean free path or increase in scattering rate of phonons, (ii) decrease in mobility of acoustic phonon-limited carriers ($\mu_P$) (see eq. A8) and (ii) increase in scattering of phonons with impurities ($P_{ph-x}$) and decrease in electron-phonon relaxation time (see eq. A14 and A15). This result is consistent with decrease in $\theta_D$, RRR and TCR with decrease in crystallite size (figure 2(a-c) and 3 (a)). Even though, electron-phonon relaxation mainly involves bulk phonons[36], decreased crystallite size, surface effects and surface phonon scattering may lead to additional phonon modes or change in their frequency that is manifested in decrease in $\theta_D$[9].

The interaction between molecules of surfactant and metal is important in the influence of surfactant. The surface-bound molecules on metal NPs can lead to change in metal work function or Fermi level, formation an interfacial dipole barrier, and charge transfer between metal and surfactant molecules that will depend upon direction and magnitude of interface dipole barrier height[4,13,31]. Therefore, different surfactants or increase in surfactant coverage may lead to change in: (i) charge carrier concentration and barrier height near the Fermi level owing to their dissimilar dielectric strengths, electron affinities and bonding strengths[4,7,13,14] and (ii) the interparticle interactions and degree of disorders, and hence (iii) the transport or tunnelling of electrons between neighbouring NPs (eq. A16). It is clear that increase in tunnelling barrier height ($\Delta E$) or interparticle separation ($\Delta x$) leads to decrease in tunnelling rate of electrons ($\Gamma$). This is analogous to decrease in differential tunnelling current around Fermi level or transmission coefficient that is inversely proportional to $\sigma(E)$, and hence will decrease the $S_d$[4]. However, as T increases, transmission coefficient increase with increase in T and hence $S_d$ increased at high T. Thus, presently observed different electrical and thermoelectric transport behaviour in NPs having different sizes and type of surfactants take place. The actual electrical and TE transport in these NPs thus involve many types of scattering, wherein GB and surfactant disorders may play a central role while spatial confinement of electrons and phonon are possibly secondary.

### 2.5 Thermoelectric power factor

Finally, it is interesting to note that the power factor $S^2/\rho$ is decreased with decrease in crystallite size in OA-TOP-induced NPs viz., Ag1, Ag2, Ag3 and Ag4 (figure 4 (c)). It is larger for Ag6 compared to Ag5 above 230 K that correspond to its smaller value of $\alpha_{e-ph}$ (figure 4(c)). The maximum value of it is found to be around 41.7 µW/m-K$^2$ at 5 K in Ag1 with its second maximum of 29.5 µW/m-K$^2$ at 5 K in Ag7 with a broad dip at around 175 K (figure 4(c), inset) that is correlated with the dip in S (figure 4 (b)) or a hump in resistivity (figure 1, inset (d)) near this temperature. Since their thermal conductivity could be relatively high as metal, their TE performance ZT is therefore likely to be relatively low without perhaps much application potentials.

### 3. Conclusion

Analysis of the measured electrical resistivity and Seebeck coefficient in T range 5 K to 300 K of well-characterized Ag nanoparticles in the size range from 15.1 nm to 33.4 nm show residual resistivity and electron-phonon coupling constant increase and Debye temperature decrease with decrease in crystallite size. Debye temperature is ~ 36 % reduced compared to bulk Ag and electron-phonon coupling constant is increased to ~ 77 % for 15.1 nm NPs compared to 31.5 nm NPs. Remarkably, usual phonon drag peak found for bulk Ag turned into dips or phonon drag minima (PDM) that gradually gets suppressed and shifted towards lower T with decrease in size in oleylamine-trioctylphosphine-induced NPs. However, it appeared at higher T in trioctylphosphine-only-induced NPs. A broad hump between 125 K to 215 K, a slope change near 270 K in resistivity and an additional dip-like feature near 172 K, in addition to the expected PDM, in S are seen in oleylamine-polyvinylpyrrolidone-made sample with different shapes of NPs. They are explained based on the spatial confinement of electrons and phonons, varying barrier heights and different charge-transfer mechanisms amongst metal NPs and surfactant/s and enhanced disorders. This study provides a fundamental understanding of thermoelectric transport of silver nanoparticles.

**APPENDIX: Sample preparation details, resistivity fitting, ratio of temperature dependent resistivity to resistivity at 5 K, temperature coefficient of resistance and some important formula of various scattering mechanisms.**

**Table A1** Sample synthesis conditions and crystallite size obtained from Scherrer relation and transmission electron microscopy (TEM).

| Sample | TOP (ml) | OA (ml) | PVP (g) | Crystallite size (nm) | TEM size (nm) |
|---|---|---|---|---|---|
| Ag1 | 1 | 10 | - | 31.5 | - |
| Ag2 | 3 | 10 | - | 28.9 | 29.1±1.2 |
| Ag3 | 5 | 10 | - | 24.7 | 25.3±0.9 |
| Ag4 | 10 | 10 | - | 15.1 | 15.9±1.1 |
| Ag5 | 8 | - | - | 29.6 | 60±2 |
| Ag6 | 12 | - | - | 24.6 | - |
| Ag7 | - | 10 | 0.25 | 33.4 | 41.7±1.2 |

The resistivity ρ of crystalline metal can be expressed using Matthiessen's rule (eq. A1).

$$\rho = \rho_0 + \rho_{e-ph}, \qquad (A1)$$

where $\rho_0$ is a T-independent part, known as a residual resistivity and dominant at very low T, mainly appeared due to scattering of electrons with GBs, surface of NPs and various defects. The T-dependent part, $\rho_{e-ph}$ arises due to electron-phonon interactions i.e. lattice contribution and can be understood using Bloch-Grüneisen (BG) function[18,28]

$$\rho_{e-ph} = \alpha_{e-ph} \left(\frac{T}{\theta_D}\right)^2 \int_0^{\frac{\theta_D}{T}} \frac{x^n dx}{(e^x-1)(1-e^{-x})}, \qquad (A2)$$

where $\alpha_{e-ph} \propto \beta_{tr}\omega_D / \omega_P^2$ is a constant that proportional to electron-phonon coupling constant $\beta_{tr}$, wherein, $\omega_D$ and $\omega_P$ are Debye frequency and plasma frequency, respectively, and $\theta_D$ is Debye temperature.

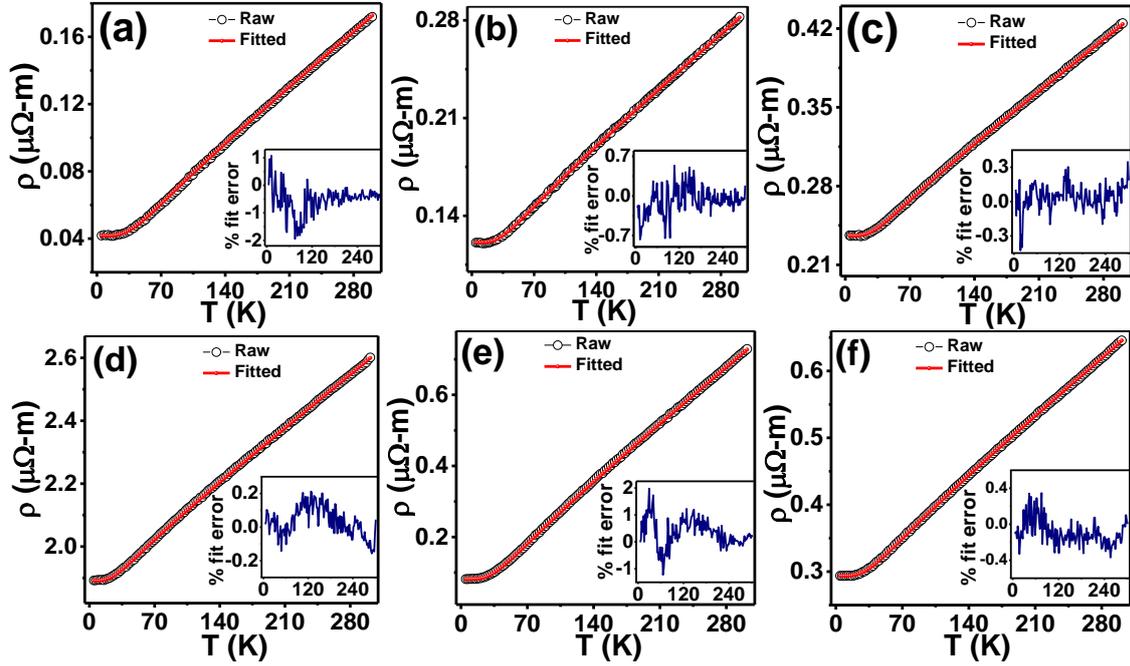

**Figure A1** Temperature dependence on resistivity of (a) Ag1, (b) Ag2, (c) Ag3, (d) Ag4, (e) Ag5 and (f) Ag6.

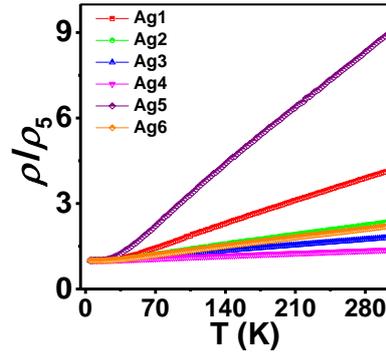

**Figure A2** Ratio of temperature dependent restively and resistivity at 5 K (a) Ag1, Ag2, Ag3, Ag4, Ag5 and Ag6.

**Table A2** Resistivity ($\rho$) at 5 K and 300 K, residual resistance ratio (RRR), Debye temperature ($\theta_D$) and electron-phonon coupling constant ($\alpha_{e-ph}$).

| Sample (Crystallite Size) | $\rho_5$ ($\mu\Omega$-m) | $\rho_{300}$ ($\mu\Omega$-m) | RRR ($\rho_{300}/\rho_5$) | $\theta_D$ (K) | $\alpha_{e-ph}$ ($\mu\Omega$-m) | Study |
|---|---|---|---|---|---|---|
| Ag1 (31.5 nm) | 0.0418 | 0.1719 | 4.112 | 220 | 0.405 | Present |
| Ag2 (28.9 nm) | 0.1207 | 0.2809 | 2.327 | 195 | 0.425 | |
| Ag3 (24.7 nm) | 0.2362 | 0.4247 | 1.798 | 180 | 0.460 | |
| Ag4 (15.1 nm) | 1.8923 | 2.5941 | 1.371 | 150 | 1.421 | |
| Ag5 (29.6 nm) | 0.0816 | 0.7244 | 8.877 | 202 | 1.770 | |
| Ag6 (24.6 nm) | 0.2937 | 0.6433 | 2.190 | 195 | 0.931 | |
| Ag7 (33.4 nm) | 0.0436 | 0.2372 | 5.440 | - | - | |
| Bulk Ag | 0.00041 | 0.0165 | 40.244 | 234 | 0.052 | |
| Polycrystalline bulk Ag | 0.0004 | 0.0164 | 41.00 | - | - | Ref[20] |
| 47 nm NPs | 0.0049 | 0.0247 | 5.040 | - | - | |
| 30 nm NPs | 0.0190 | 0.0398 | 2.094 | - | - | |
| 20 nm NPs | 0.0269 | 0.0502 | 1.866 | - | - | |

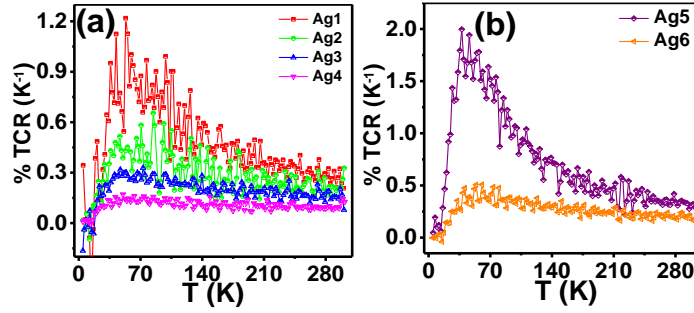

**Figure A3** Temperature coefficient of resistance CR of (a) Ag1, Ag2, Ag3 and Ag4, (b) Ag5 and Ag6 as function of T.

The diffusion contribution in S ($S_d$) can be described using Mott's formula[10]

$$S_d = -\frac{\pi^2 k_B^2 T}{3e}\left[\frac{1}{\sigma}\frac{d\sigma(E)}{dE}\right]_{E=E_f} \quad (A3)$$

Energy dependent electrical conductivity can be given by[10]

$$\sigma(E) = \frac{e^2 l(E) a(E)}{6\pi^2 h} = \frac{ne^2 \tau(E)}{m} \quad (A4)$$

where energy dependent mean free path of electron $l(E) = \tau(E)v(E)$. $\tau(E)$, $v(E)$ and $a(E)$ are energy-dependent relaxation time, electron velocity and area of the surface in k-space[37].

The $S_d$ in the free-electron approximation can be given by[37]

$$S_d = \frac{\pi^2 k_B^2 T}{3eE_F} \quad (A5)$$

for electron-residual impurity scattering at low T and

$$S_d = \frac{\pi^2 k_B^2 T}{eE_F} \quad (A6)$$

for electron-phonon scattering at high T.

The S for metals and degenerate semiconductors with parabolic band and energy-independent charge carrier scattering can be given by[44]

$$S_d = \frac{8\pi^2 k_B^2 T}{3eh^2} m \left(\frac{\pi}{3n}\right)^{2/3}, \quad (A7)$$

where n, m and $k_B$ are charge carrier concentration, mass of electron and Boltzmann constant.

Herring's formula for phonon drag contribution for 2D electron gas can be given as[11]

$$S_g = \frac{v_P \tau_P^2}{\mu_P T}, \quad (A8)$$

where $v_P$, $\tau_P$ and $\mu_P$ are phonon group velocity, phonon relaxation time involve in phonon drag and acoustic phonon-limited mobility of electrons.

The $S_d$ and $S_g$ in the presence of different kinds of scatterings are on the other hand written as[37]

$$S_d^{e-ph} \propto \frac{\tau^2 v_F T^4}{E_F u_l^2}, \quad (A9)$$

for pure electron-phonon interaction contribution to $S_d$,

$$S_d^{e-ph-imp} \propto \frac{\tau T^3}{E_F u_l}, \quad (A10)$$

for contribution of inelastic scattering of electrons due to impurities with the emission or absorption of a phonon to $S_d$,

$$S_g^{e-ph} \propto -\frac{e\tau T^3}{mu_l^3}, \qquad (A11)$$

as contribution of thermopower to $S_g$ due to phonon drag,

$$S_g^{e-imp} \propto \frac{u_l p_F}{E_F \tau T}, \qquad (A12)$$

as contribution of inelastic scattering of electrons on impurity to $S_g$.

The phonon drag contribution to the total S for impure metal can be given by

$$S_g = \frac{2\pi^2 T^3}{45 n_0 e u_l^3}\left(1 - 1.06 \frac{h u_l p_F}{E_F \tau T}\right), \qquad (A13)$$

where, $v_F$, $p_F$ and $u_l$ are the Fermi velocity, Fermi momentum and wave vector of longitudinal phonon involved in this scattering. The $\tau$ is relaxation time for corresponding scattering that will not be different for each scattering mechanism. Here role of transverse phonons is not considered.

The relation between $S_g$ and probability of electron-phonon interaction ($P_{e-ph}$) and probability of scattering of phonons with any potential except electron ($P_{ph-x}$) can be written as[45]

$$S_g \propto \frac{P_{e-ph}}{P_{ph-x} + P_{e-ph}} \qquad (A14)$$

and

$$P_{e-ph} \sim \frac{P_{e-ph}^{bulk}}{\exp(\eta \omega_{min}/KT) - 1}, \qquad (A15)$$

where $P_{e-ph}^{bulk}$ and $\omega_{min} = \frac{\pi v}{d}$, wherein $v$ and d are sound velocity and diameter of the particle.

The quantum mechanical coupling between the energy levels can be expressed in terms of the coupling energy ≈ hΓ, where h is Planck's constant and Γ is the tunnelling rate between two orbitals of nanocrystal neighbours. The tunnelling rate can be approximated as[7]

$$\Gamma \approx \exp\left[-4\left(\frac{2m^* \pi^2 \Delta E}{h^2}\right)^{1/2} \Delta x\right], \qquad (A16)$$

where, Γ, $m^*$, $\Delta E$ and $\Delta x$ are tunnelling rate, effective mass of electron, barrier height for tunnelling and shortest interparticle separation between the nearest neighbour NPs, respectively.